\documentclass[11pt]{article}
\pdfoutput=1
\usepackage{dcolumn}
\usepackage{bm}

\usepackage{graphicx}
\usepackage{amssymb,amsmath}
\usepackage{multirow}
\usepackage{cite,color,url}
\usepackage[colorlinks=true
,urlcolor=blue
,anchorcolor=blue
,citecolor=blue
,filecolor=blue
,linkcolor=blue
,menucolor=blue
,linktocpage=true
,pdfproducer=medialab
,pdfa=true
]{hyperref}

\usepackage{slashed}
\usepackage{epsfig,psfrag,rotating,soul}
\usepackage{rotfloat}

\evensidemargin \oddsidemargin
\allowdisplaybreaks

\let\OLDthebibliography\thebibliography
\renewcommand\thebibliography[1]{
  \OLDthebibliography{#1}
  \setlength{\parskip}{0pt}
  \setlength{\itemsep}{0pt plus 0.3ex}
}

\begin{document}
\thispagestyle{empty}

\def\thefootnote{\fnsymbol{footnote}}

\begin{flushright}
IFT-UAM/CSIC-18-86\\
\end{flushright}

\vspace*{1cm}

\begin{center}

\begin{Large}
\textbf{\textsc{Constraining R-axion models through dijet searches at the LHC}}
\end{Large}

\vspace{1cm}

{\sc
Ernesto~Arganda$^{1, 2}$%
\footnote{{\tt \href{mailto:ernesto.arganda@fisica.unlp.edu.ar}{ernesto.arganda@fisica.unlp.edu.ar}}}%
, An\'{\i}bal D.~Medina$^{1}$%
\footnote{{\tt \href{mailto:anibal.medina@fisica.unlp.edu.ar}{anibal.medina@fisica.unlp.edu.ar}}}%
, Nicol\'as~I.~Mileo$^{1}$%
\footnote{{\tt \href{mailto:mileo@fisica.unlp.edu.ar}{mileo@fisica.unlp.edu.ar}}}%
, Roberto A.~Morales$^{1}$%
\footnote{{\tt \href{roberto.morales@fisica.unlp.edu.ar}{roberto.morales@fisica.unlp.edu.ar}}}%
and Alejandro Szynkman$^{1}$%
\footnote{{\tt \href{mailto:szynkman@fisica.unlp.edu.ar}{szynkman@fisica.unlp.edu.ar}}}%
}

\vspace*{.7cm}

{\sl
$^1$IFLP, CONICET - Dpto. de F\'{\i}sica, Universidad Nacional de La Plata, \\ 
C.C. 67, 1900 La Plata, Argentina

\vspace*{0.1cm}

$^2$Instituto de F\'{\i}sica Te\'orica UAM/CSIC, \\ Calle Nicol\'as
Cabrera 13-15, Cantoblanco E-28049 Madrid, Spain
}

\end{center}

\vspace{0.1cm}

\begin{abstract}
\noindent
The search at hadron colliders for new massive  resonances of a few 100 GeVs that couple effectively to colored states is an extremely challenging issue, due principally to the presence of large QCD multijet backgrounds at this energy, rendering the searches at the LHC particularly difficult. Recently, it was realized that these large backgrounds could be overcome by demanding one high-$p_T$ jet from initial-state radiation (ISR) and by means of novel jet-reconstruction techniques through which the resulting hadronized products of the massive resonances are reconstructed as a fat-jet, a unique large-radius jet. The ATLAS and CMS Collaborations have recently reported searches for the experimental signature of a single fat-jet in association with an ISR jet. Models of dynamical supersymmetry breaking with an spontaneously broken R-symmetry give rise to the appearance of a pseudo-Nambu-Goldstone boson called the R-axion, which naturally tends to be light. In the parameter space regions where the anomalous R-axion coupling to gluons is boosted, these models can be tested against these new LHC dijet searches. Taking into account the CMS search, we apply the $q_\mu$ statistical method to the signal events against the background-only expectation and obtain the 95\% C.L. exclusion limits on the most relevant model parameters for a particular messenger sector, namely, the R-axion mass $m_a$, the decay constant $f_a$, and the number of color messengers $N$, being these limits suitable to be applied to more general models with axion-like particles.
\end{abstract}

\def\thefootnote{\arabic{footnote}}
\setcounter{page}{0}
\setcounter{footnote}{0}

\newpage

\section{Introduction}
\label{intro}

Models of dynamical supersymmetry (SUSY) breaking provide,  via dimensional transmutation $m_\text{weak}\sim e^{-\mathcal{O}(1) 8\pi^2/g^2}M_\text{Planck}$, an elegant solution to the  problem of hierarchies between the a priori unrelated scales of  weak ($m_\text{weak}\approx \mathcal{O}(100)$ GeV) and gravitational interactions ($M_\text{Planck}\approx \mathcal{O}(10^{19})$ GeV). It was shown in~\cite{Nelson:1993nf} that in a generic class of dynamical  models in which  SUSY is broken via F-terms, an spontaneously broken R-symmetry is a sufficient condition for dynamical SUSY breaking, thus making the study of SUSY models with a $U(1)_R$ symmetry very appealing.  The spontaneous breaking  of the global R-symmetry leads to the appearance of a Nambu-Goldstone boson called the R-axion. There have been several studies about the nature and phenomenology of the R-axion~\cite{Bagger:1994hh,Goh:2008xz,Bellazzini:2017neg}. The cancellation of the cosmological constant  provided by the tuning of a constant term in the supergravity (SUGRA) superpotential  leads to an explicit breaking of the $U(1)_R$ and thus to  an unavoidable SUGRA contribution to the R-axion mass, making the R-axion a pseudo-Nambu-Goldstone boson (pNGB). There can be however other explicit sources of $U(1)_R$ breaking in the hidden sector that add up to the SUGRA contribution, but keep it light with respect to the other SUSY particles of the theory, respecting the pNGB nature and therefore possibly making it the first sign of SUSY. The couplings of the R-axion to other particles depend on the $U(1)_R$ charge assignments one makes on the different sectors of the theory (messenger and visible sectors). Nevertheless, if the sources of explicit $U(1)_R$ breaking are kept small, the R-axion couplings still display its Goldstone nature and thus are suppressed by its decay constant $f_a$, which is related to the scale of $U(1)_R$ breaking.

In~\cite{Bellazzini:2017neg} a phenomenological study of the R-axion was made where its decay constant and mass were kept as free parameters. The study focused on the possible collider signatures, and assumed that there is a strongly coupled hidden sector where SUSY is broken and from which SUSY breaking is transmitted to the visible sector {\it \`a la} gauge mediation. The $U(1)_R$ is realized non-linearly by the R-axion and SUSY is realized non-linearly in the constrained superfield formalism to capture the low-energy behaviour, providing the interactions between the R-axion and the effective Gravitino Lagrangian. In particular, the coupling of the R-axion to gravitino pairs\footnote{The coupling is dominated by the Goldstinos which are longitudinal modes of the gravitino.} can be quite sizable in certain regions of parameter space. Given that Majorana gaugino masses break explicitly the $U(1)_R$, one expects a SUSY spectrum where gaugino-like neutralinos (and possibly Higgsinos) tend to be lighter than the other soft-breaking masses. Thus it is possible to obtain a somewhat natural spectrum at low energies,  where the only light particles in the spectrum related to SUSY are the gravitino, the R-axion and possibly the neutralinos. 

In this work we propose to test a class of R-axion models against the dijet searches recorded recently by the CMS Collaboration~\cite{Sirunyan:2017nvi} at the Large Hadron Collider (LHC) with a center-of-mass energy of $\sqrt{s} =$ 13 TeV and a total integrated luminosity of $\mathcal{L}=35.9$ fb$^{-1}$. The experimental signature consists of a single massive large-radius jet in association with a jet from initial-state radiation, and the collaboration reports a slight deviation with respect the background-only expectation of 2.9$\sigma$ of local significance and 2.2$\sigma$ of global significance. The ATLAS Collaboration has performed a more recent search~\cite{Aaboud:2018zba} for new resonances identified as massive large-radius jets consistent with a particle decaying into quark pairs, obtaining lower values for the local and global significances of the observed deviations above background. We will make use of some of these experimental results to probe the mentioned models of dynamical SUSY breaking in which the anomalous R-axion coupling to gluons is boosted, imposing constraints on the most relevant model parameters, namely, the R-axion mass $m_a$, the decay constant $f_a$, and the number of color messengers $N$. In addition, these limits should not be difficult to generalized to more generic axion-like particle (ALP) models that posses a similar coupling structure to the R-axion models exhibited here. Analyses of this type have been carried out in~\cite{Mariotti:2017vtv} recasting the limits imposed on $Z^\prime$ models reported by CMS in~\cite{Sirunyan:2017nvi}.

This paper is organized as follows: In Section~\ref{sec:theory} we provide a brief theoretical review on the R-axion and its most relevant interactions. In Section~\ref{sec:CMS} we comment on the recent dijet searches performed by the ATLAS and CMS collaborations, while Section~\ref{sec:interpretation} is devoted to determine the constraints imposed by these experimental searches on the R-axion model. Finally, we conclude in Section~\ref{sec:conclusions}.


\section{Brief theoretical review on R-axion}
\label{sec:theory}
We refer the reader to~\cite{Goh:2008xz} and~\cite{Bellazzini:2017neg}  for a more thorough  description of the framework. We define R-symmetry as the largest subgroup of the automorphism group of the supersymmetry algebra that commutes with the Lorentz group. We focus on the Minimal Supersymmetric extension of the SM (MSSM) and as we mentioned in the Introduction, we  expect a natural spectrum where squarks and sleptons of all families are decoupled from the low-energy effective theory. In fact, we consider that the only SUSY particles that remain in the low-energy spectrum are the R-axion and possibly the gravitino\footnote{The gravitino tends to be light in the case that the F-term responsible for SUSY breaking is small. However, it has no impact on the phenomenology we are interested in.}. The relevant interactions of the R-axion for the phenomenology we wish to describe are given by its coupling to the MSSM gauge sector and in particular its anomalous coupling to gluons and photons.  We assume that there exist two classes of messenger fields that transmit SUSY breaking to the visible sector. One is a {\it single} $5+\bar{5}$ of $SU(5)$ which contributes to the masses of gauginos and sfermions in the usual general gauge-mediated way. The other class consists of $N$-copies of messengers $q+\bar{q}$ which are $3+\bar{3}$ under $SU(3)_c$ and singlets under $SU(2)_L$ and $U(1)_Y$ weak gauge groups. This messenger sector upsets the unification of gauge couplings at high energies but that is something we are not concerned with in this work. It also implies that gluinos and squarks naturally tend to be heavier than weak gauginos and sleptons. This kind of spectrum is where the current searches at the LHC seem to be leading us towards given the lack of evidence of sparticles, in particular colored ones like gluinos and squarks  which should be easily produced at a hadron collider. Another reason to consider this messenger sector lies in the fact that current diphoton resonant searches~\cite{Aad:2014ioa,Khachatryan:2016yec} already put strong constraints in the R-axion parameter space if one decides to consider the somewhat standard choice of $N$ copies of $5+\bar{5}$  under $SU(5)$~\cite{Bellazzini:2017neg}. Allowing to split the messenger sector as we do implies that R-axions now become basically insensitive to diphoton searches but nonetheless, as we will show in later sections, they can still be probed via the novel fat-jet techniques  that profit from the anomalous coupling to gluons. Following~\cite{Goh:2008xz}, we can write the relevant couplings of the R-axion to massless gauge bosons as
\begin{eqnarray}
\Delta\mathcal{L}^{{\rm eff}}_{agg}&=&\frac{g^2_{s}}{32\pi^2}\frac{a}{f_a}\times\left[-N-1+3-2\times\frac{\cos^2\beta}{2}\right]G_{\mu\nu}\tilde{G}^{\mu\nu}\label{couplagg} \,,\\
\Delta\mathcal{L}^{{\rm eff}}_{a\gamma\gamma}&=&\frac{e^2}{32\pi^2}\frac{a}{f_a}\times\left[-2+2-3\times\frac{4}{9}\times2 \cos^2\beta\right]F_{\mu\nu}\tilde{F}^{\mu\nu} \,, \label{couplagamgam}
\end{eqnarray}
where $G_{\mu\nu}$ and $F_{\mu\nu}$ are the gluon and photon field strengths, $\tilde{G}_{\mu\nu}=\epsilon_{\mu\nu\rho\sigma}G^{\rho\sigma}/2$  is the dual gluon field strength and similarly for the photon, and  we have explicitly separated the contribution of the single $5+\bar{5}$ (the $-1$ and $-2$ in the couplings to gluons and photons, respectively)  from that of the $N$-copies of $3+\bar{3}$ of $SU(3)_c$ which only enter in the coupling to gluons. Furthermore, we have assumed that all of the gauginos are heavy enough such that their loop contribution can be taken as an anomalous contribution as well (the $+3$ in the coupling to gluons and the $+2$ in the coupling to photons). The only SM fermion contribution that we are taking into account (given that we plan to   consider an R-axion with a mass $m_a\approx \mathcal{O}(100)$ GeV) comes from the top quark, which enters in an indirect way given that we assume no direct coupling of the R-axion to fermions. In fact, we consider an R-symmetry consistent with the $\mu$-term and Yukawa interactions, $r_{H_u}+r_{H_d}=2$, $r_{H_u}+r_{Q}+r_{U}=2$, $r_{H_d}+r_{Q}+r_{D}=2$, $r_{H_d}+r_{L}+r_{E}=2$, and in which the R-charges of $H_u$ and $H_d$ are fixed by the condition that the Goldstone boson associated to the $Z$-gauge boson is invariant under $U(1)_R$, ( $r_{H_u}=2\cos^2\beta$,   $r_{H_d}=2\sin^2\beta$ with $\tan\beta=\langle H_u\rangle/ \langle H_d\rangle$). It is clear then that there is no direct coupling of the R-axion to fermions.  Instead what happens is that the R-axion mixes with the CP-odd Higgs via the $B_{\mu}$ term in the Higgs potential, and this dictates the way in which the R-axion effectively couples to fermions, via the mixing and the CP-odd Higgs coupling to fermions. We will follow ~\cite{Goh:2008xz}, and assume that the $B_{\mu}$ term is generated via Renormalization Group running from gaugino-loops. In that case, the effective coupling of the R-axion to SM fermions takes the form,
\begin{equation}
\Delta\mathcal{L}^{{\rm eff}}_{af\bar{f}}=i \frac{a}{f_a}(\cos^2\beta m_u \bar{u}\gamma_5 u+\sin^2\beta m_d \bar{d}\gamma_5 d+\sin^2\beta m_l \bar{l}\gamma_5 l) \,. \label{couplaff}
\end{equation}
 Thus we understand the appearance of the top quark in the anomalous coupling of the R-axion to gluons and photons as it enters in the top quark fermion loop triangle diagram. For example, in the anomalous contribution to gluons, the factor of $1/2$ stems from the trace over the $SU(3)_c$ generators and the additional factor of 2 comes from the different directions in the fermion loop.  In the anomalous coupling to photons, there is a color factor  $N_c=3$, the top-quark electromagnetic charge squared and once again a factor of two from the two possible directions of the fermion loop. The factor $\cos^2\beta/f_a$ which is common in both diagrams comes from the effective coupling to up-type fermions, Eq.~(\ref{couplaff}).


We would like to stress that our results could be easily accommodated to any other different set of R-charges, for example one in which the $\mu$-term carries non-vanishing R-charge when generated  by the hidden sector dynamics, which would then modify $r_{H_u}+r_{H_d}$ and with it all the other choices for the fermion R-charges. Similarly, as was done in ~\cite{Bellazzini:2017neg}, one could assume that the hidden dynamics is responsible for the generation of the $B_{\mu}$ term, in which case the effective coupling of the R-axion to fermions would be modified. Given that we plan to consider values of $\tan\beta\gg 1$ rendering the top-quark contribution marginal, and that we also consider R-axion masses such that the light fermion contributions to the anomalies are negligible, we expect that this latter modification on the mixing to be completely negligible. 


Therefore, the phenomenology in which we are interested in is described by the anomalous coupling to gluons and photons in Eqs.~(\ref{couplagg}) and~(\ref{couplagamgam}), as well as the coupling to fermions in Eq.~(\ref{couplaff}). The relevant parameters are the number of color messengers, $N$, the R-axion mass, $m_a$, the decay constant, $f_a$, and $\tan\beta$. 
Throughout all the analysis performed in this paper we will vary $N$ and $m_a$ in the ranges [1,10] and [45 GeV, 305 GeV], respectively, and consider $f_a=1$ TeV as a benchmark. Finally, we will fix the value of $\tan\beta$ to $10$, though our results do not depend on this particular choice. In fact, we expect that our conclusions will not change for $4 \lesssim \tan\beta\lesssim 30$.

\section{Low mass dijet resonances at the LHC}
\label{sec:CMS}

The search at hadron colliders for low-lying-massive resonances that couple to colored states is extremely challenging due to the presence of large SM backgrounds, principally the so-called QCD multijet background. This issue can be faced by demanding one high-$p_T$ jet from initial-state radiation, which allows to satisfy the energy trigger requirements. Under these conditions, the CMS Collaboration has recently considered this class of search with an enough high $p_T$ for the new resonances~\cite{Sirunyan:2017nvi}, in such a way that the resulting hadronized products are reconstructed as an unique large-radius jet, usually called fat-jet. CMS has looked for narrow resonance peaks in the continuous distributions of fat-jet masses, taking advantage of a jet-mass and $p_T$-decorrelated substructure variable which keeps the jet-mass distribution shapes. This novel variable is the jet-mass distribution groomed with the soft-drop algorithm~\cite{Dasgupta:2013ihk,Larkoski:2014wba}, $m_\text{SD}$, which reduces the distribution of jet masses of the QCD background arising from soft ISR gluons whilst keeps practically unchanged the jet masses coming from the new massive resonances and $W/Z$ backgrounds. The fat-jet candidate in each event is the most energetic jet that fulfills the requirement to be an AK8 jet (reconstructed using the anti-$k_\text{T}$ algorithm with a radius parameter of 0.8) with $p_T >$ 500 GeV and $|\eta| <$ 2.5. The distribution for the collected data, corresponding to $\mathcal{L}=35.9$ fb$^{-1}$ at $\sqrt{s} =$ 13 TeV, and the simulated SM background of the leading-jet soft-drop mass $m_\text{SD}$ is displayed in Figure 1 of~\cite{Sirunyan:2017nvi}, while in Figure 6 the $m_\text{SD}$ distribution is shown for data and background in five different $p_T$ ranges (500, 600, 700, 800, 900, and 1000 GeV) considered by CMS for a signal consisting of a leptophobic vector boson $Z^\prime$ with a mass of 135 GeV. After computing the upper limits on the production cross section by means of the CL$_S$ method~\cite{Junk:1999kv,Read:2002hq,Cowan:2010js}, a maximum local (global) excess of 2.9 (2.2) standard deviations is observed at $m_\text{SD} \simeq$ 115 GeV. 

More recently, the ATLAS Collaboration has reported a similar search~\cite{Aaboud:2018zba}, corresponding to $\mathcal{L}=36.1$ fb$^{-1}$ at also $\sqrt{s} =$ 13 TeV. In this case, ATLAS uses two non-exclusive jet categories defined by the radius parameter $R$ of the anti-$k_\text{T}$ algorithm: {\it large}-$R$ jets ($J$) with $R =$ 1.0 and $|\eta| <$ 2.0 and {\it narrow} jets ($j$) with $R =$ 0.4 and $|\eta| <$ 2.4. The event selection implies the requirement of having at least one large-$R$ jet with $p_T^J >$ 450 GeV (the resonance candidate) and at least one narrow jet (the ISR jet) with $p_T^j >$ 420 GeV and azimuthal angular separation of $\Delta\phi > \pi/2$ with respect to the resonance candidate. The large-$R$ jet mass distributions of data and background, not displayed in $p_T$ bins as in the CMS analysis, are presented in Figure 3 and 4 of~\cite{Aaboud:2018zba}, where it can be seen that the estimated background contributions reproduce very well the observed distributions. After a signal-plus-background fit to the large-$R$ jet mass distribution with a $Z^\prime$ model assumption, a local (global) excess of 2.5$\sigma$ (1.1$\sigma$) is observed for masses around 150 GeV.


The R-axion model described in Section~\ref{sec:theory} can actually lead to a final state signature with a fat-jet plus an ISR jet via the process $pp\to a(\to jj)j$. Although both the ATLAS and CMS searches can lead to constraints on the R-axion model, from now on we will only focus on the latter. On the one hand, the mass range explored by the CMS search (50-300 GeV) is wider than the one considered in the ATLAS analysis (100-220 GeV). On the other hand, the reconstruction of the two non-exclusive jet categories used by ATLAS is more difficult to deal with by means of a fast simulation of the detector response, and this would make our recasting of the ATLAS results less reliable than in the case of CMS. In addition, in the mass range where ATLAS and CMS analyses overlap, the exclusion limits on the signal cross sections are of the same order and therefore, we do not expect that our results would be significantly modified by the inclusion of the ATLAS analysis.

\begin{figure}
\begin{center}
\centering
\includegraphics[scale=0.75]{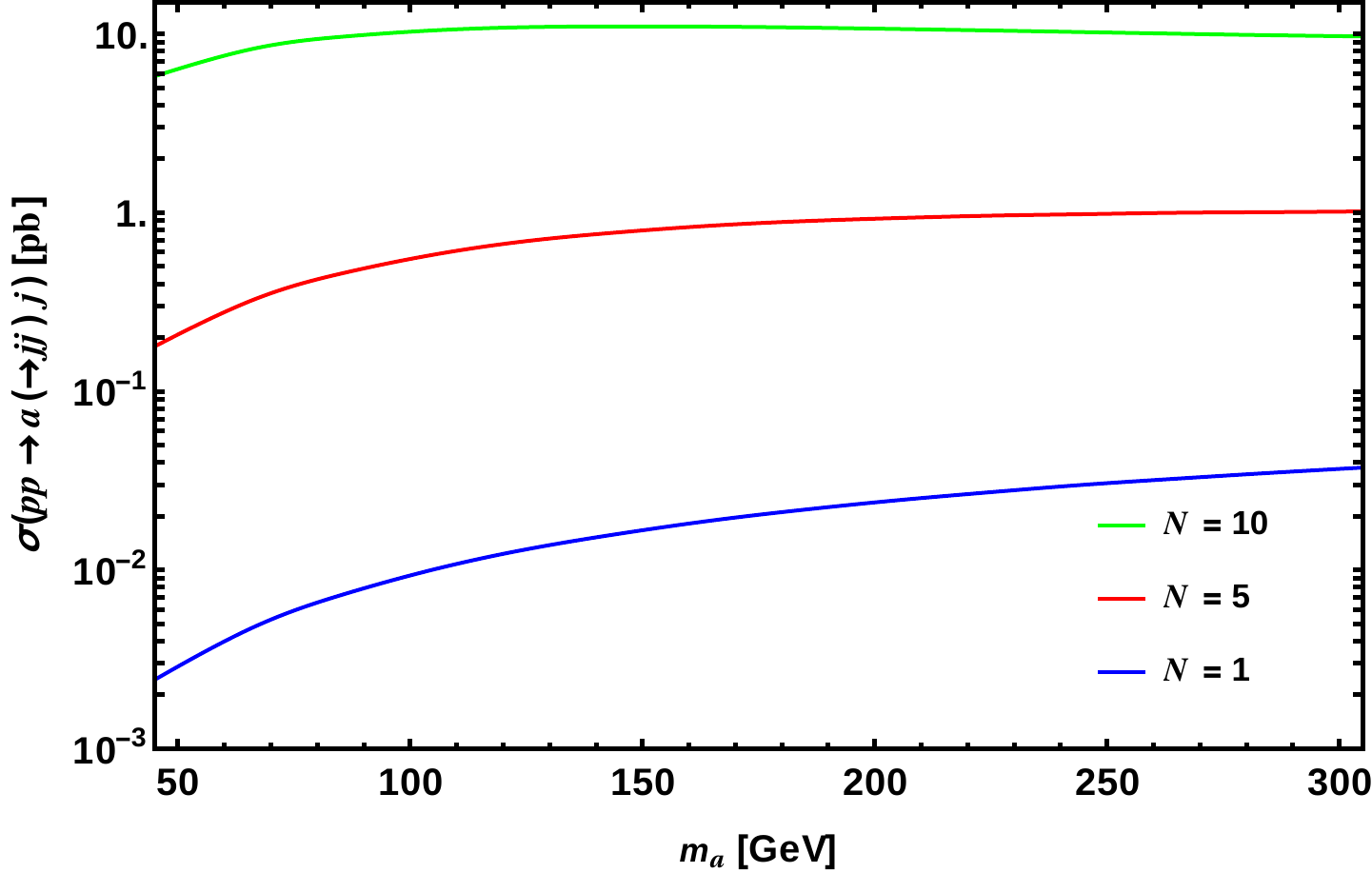}
\caption{Signal cross section as a function of $m_a$ at 13 TeV and $f_a=1$ TeV for $N=1,\,5,\,10$, computed with {\tt MadGraph 5} with matching up to one extra jet after imposing $p_T^\text{leading-jet} >$ 300 GeV at the generation level. The cross section includes an R-axion mass dependent $K$-factor correction~\cite{Ahmed:2016otz}.}
\label{signalxs}
\end{center}
\end{figure}

In Fig.~\ref{signalxs} we show the signal cross section $\sigma (pp \to a(\to jj)j) =\sigma (pp \to aj) \times$BR$(a \to jj)$ at 13 TeV for three values of $N$ computed with {\tt MadGraph 5}~\cite{Alwall:2014hca} with matching up to one extra jet after imposing a cut on the transverse momentum of the leading jet $p_T^\text{leading-jet} >$ 300 GeV at the generator level\footnote{This cut is imposed in order to make a more efficient generation while considering a possible fraction of events generated with $p_T^\text{leading-jet}<$ 500 GeV passing the cut $p_T^\text{leading-jet}>$ 500 GeV applied in the CMS analysis at the reconstruction level.}. In addition, an R-axion mass dependent $K$-factor, which includes full NNLO, approximate N$^3$LO and a threshold resummation at N$^3$LL'~\cite{Ahmed:2016otz}, has been applied to the signal cross section according to Fig. 7 of that reference. Note that this figure provides the $K$-factor down to masses of 100 GeV and then, in order to cover the region between 45 GeV and 100 GeV, we have performed an extrapolation by adding a $K$-factor of 3.7 for $m_a$ = 45 GeV in consonance with~\cite{Mariotti:2017vtv}. As can be seen from Fig.~\ref{signalxs}, within the considered values for $N$ and $m_a$ the cross section spans the range of $\sim 10^{-3}-10$ pb. A comment is in order about the particular case in which $N=2$. From Eq.~(\ref{couplagg}) we see that for this value there is a cancellation that results in an anomalous coupling to gluons proportional to $r_t=\cos^2\beta \sim 10^{-2}$ (since we use $\tan \beta = 10$). This suppression in the coupling gives rise to a signal cross section many orders of magnitude smaller than the one obtained for the remaining values of $N$. 
 
\section{Results}
\label{sec:interpretation}

\begin{figure}
\begin{center}
\centering
\includegraphics[scale=0.65]{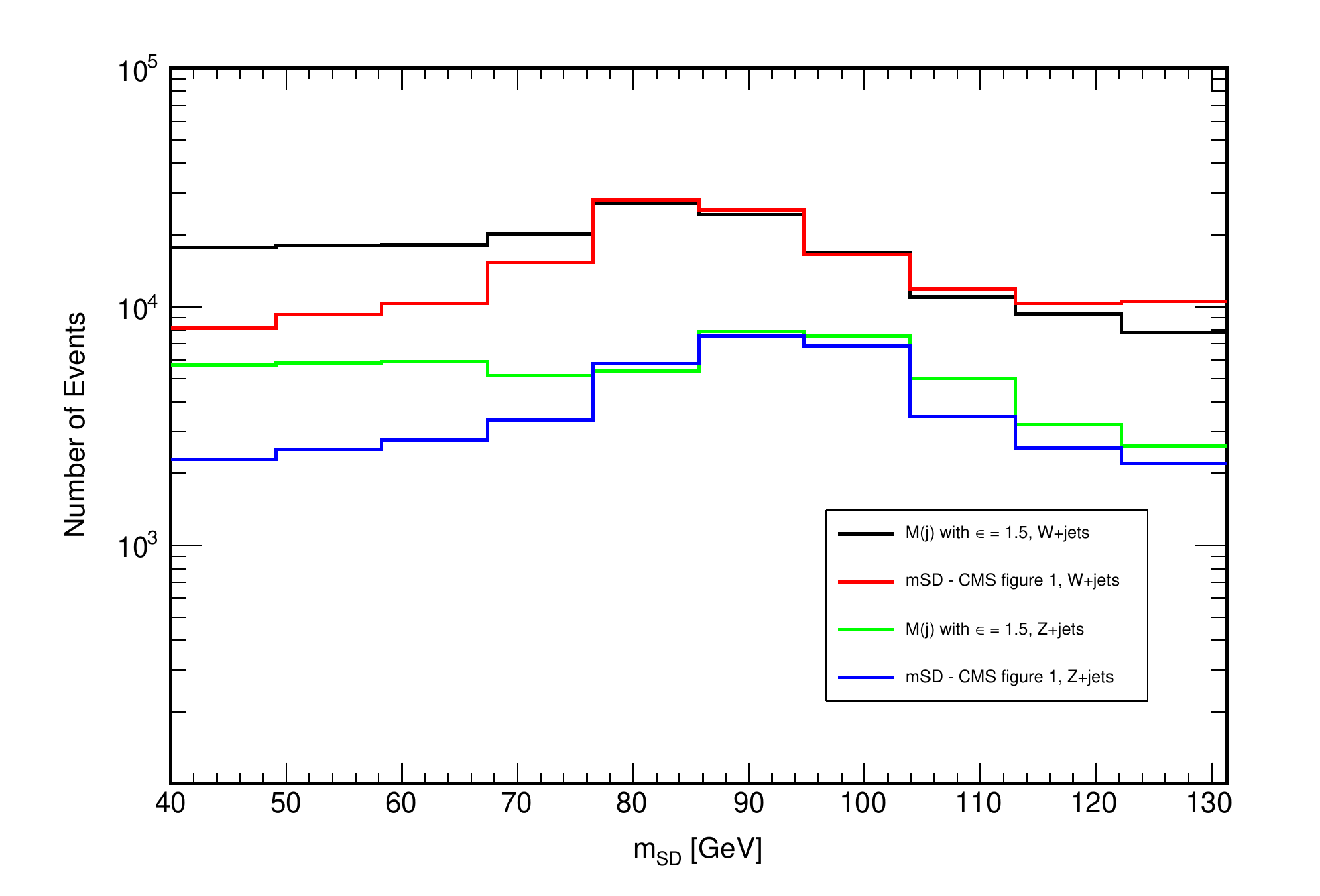}
\caption{Comparison of the $m_\text{SD}$ distribution and the invariant mass distribution corrected by a factor of $1.5$ for the SM processes $W$+jets and $Z$+jets.}
\label{figcomparison}
\end{center}
\end{figure}

In order to determine the constraints imposed by the CMS searches on the R-axion model, it is necessary to obtain the $m_\text{SD}$ distribution corresponding to our signal. However, the soft-drop algorithm required to reconstruct this variable is out of the scope of this work. Instead we used directly the invariant mass distribution of the leading jet corrected by a factor $\epsilon$ obtained from the comparison between the number of events at the peak of the $m_\text{SD}$ and the invariant mass distributions corresponding to the $W$+jets background. Here we are assuming that the resulting correction factor can also be used to reproduce the resonance peak in the $m_\text{SD}$ distribution of the signal. As it is stated in~\cite{Sirunyan:2017nvi}, the jet mass for merged $W \rightarrow q \bar q$ (as well as for $Z \rightarrow q \bar q$ and $Z' \rightarrow q \bar q$) jets is mostly unchanged by the soft-drop grooming since it arises from the kinematics of the decay process as opposed to QCD background jets. Therefore, expecting the same to happen with $a \rightarrow j j$, it is not unlikely to assume a similar correction factor for our signal.


We generated  $4.5\times 10^5$ events of the $W$+jets background at LO using {\tt MadGraph 5}~\cite{Alwall:2014hca}, with the $W$ boson decaying hadronically. The parton shower and hadronization were carried out with {\tt PYTHIA 8}~\cite{Sjostrand:2014zea} and the detector response was implemented with {\tt Delphes 3}~\cite{deFavereau:2013fsa}. Also, we matched the sample up to three additional jets. Regarding the jet reconstruction, we use the anti-$k_\text{T}$ algorithm with the $R$ parameter set to 0.8. Hence, all the jets in the event are reconstructed as fat-jets and we associate the fat-jet candidate with the most energetic AK8 jet.


In order to obtain the correction factor mentioned above, we plot with {\tt MadAnalysis5} \cite{Conte:2012fm,Conte:2014zja,Dumont:2014tja,Conte:2018vmg} the invariant mass distribution of the leading jet after applying the cuts $p_T>500$ GeV and $|\eta|<2.5$ used in~\cite{Sirunyan:2017nvi}. From this simulated distribution we found that the number of events at the peak exceeds that of the $m_\text{SD}$ distribution by a factor of $\epsilon=1.5$. In order to test the robustness of this value, we also performed the estimation of the correction factor by using the $Z$+jets background. In this case we obtain a value $\sim 1.6$, whose difference with the factor derived from the $W$+jets background is negligible in terms of its impact on the exclusion limits. Based on this check, we used the same correction factor regardless of the resonance mass. In Fig.~\ref{figcomparison} we show the $m_\text{SD}$ distribution for the $W$+jets and the $Z$+jets backgrounds provided by CMS in~\cite{Sirunyan:2017nvi} along with our estimation through the invariant mass distribution and a correction factor of $1.5$ in the vicinity of the resonance peak. It is important to stress that even when ten bins of $m_\text{SD}$ are displayed in this figure, we concentrate, when estimating the correction factor, on the two bins with the highest number of events. By looking at Fig.~\ref{figcomparison} one may think that a correction factor $\sim 3$ would be more appropriate than $1.5$ for the region below the resonance peak. However, even overestimating this possible change in the correction factor outside the resonance peak and using the value $3$ instead of $1.5$ to correct the whole distribution, the exclusion limits are slightly weakened in half of the considered mass range. Finally, potential variations of the correction factor in the high invariant mass region above the resonance peak do not affect our results since the invariant mass distribution of the signal falls abruptly there. 



Regarding the simulation of the signal, we use the same setup as for the $W/Z$+jets backgrounds and consider values of $N$ in the range $[1,10]$ and resonance masses between $45$ GeV and $305$ GeV (as in~\cite{Sirunyan:2017nvi}) in steps of $20$ GeV. For each point of this grid we generate $5\times10^4$ events and build the leading jet invariant mass distribution in the five different $p_T$ ranges considered in the CMS analysis (see Figure 6 of~\cite{Sirunyan:2017nvi}). These invariant mass distributions are in turn mapped into $m_\text{SD}$ distributions  by means of the correction factor $\epsilon=1.5$. The $m_\text{SD}$ distributions corresponding to the different backgrounds were taken from the Figure 6 of \cite{Sirunyan:2017nvi}. In addition to the correction factor, we need to apply the acceptance corresponding to the cut $N^{1,\text{DDT}}_2<0$ that is included in the CMS analysis to discriminate two-prong signal jets from multijet QCD background jets preserving the shape of the soft-drop jet mass distribution. From Figure 3 of \cite{Sirunyan:2017nvi}, the acceptance is found to be $\sim 0.3$ both for the $W$+jets background and for the $Z'$ model tested in that reference.

\begin{figure}[t!]
\begin{center}
\hspace*{-5mm}
\begin{tabular}{cc}
\includegraphics[width=83mm]{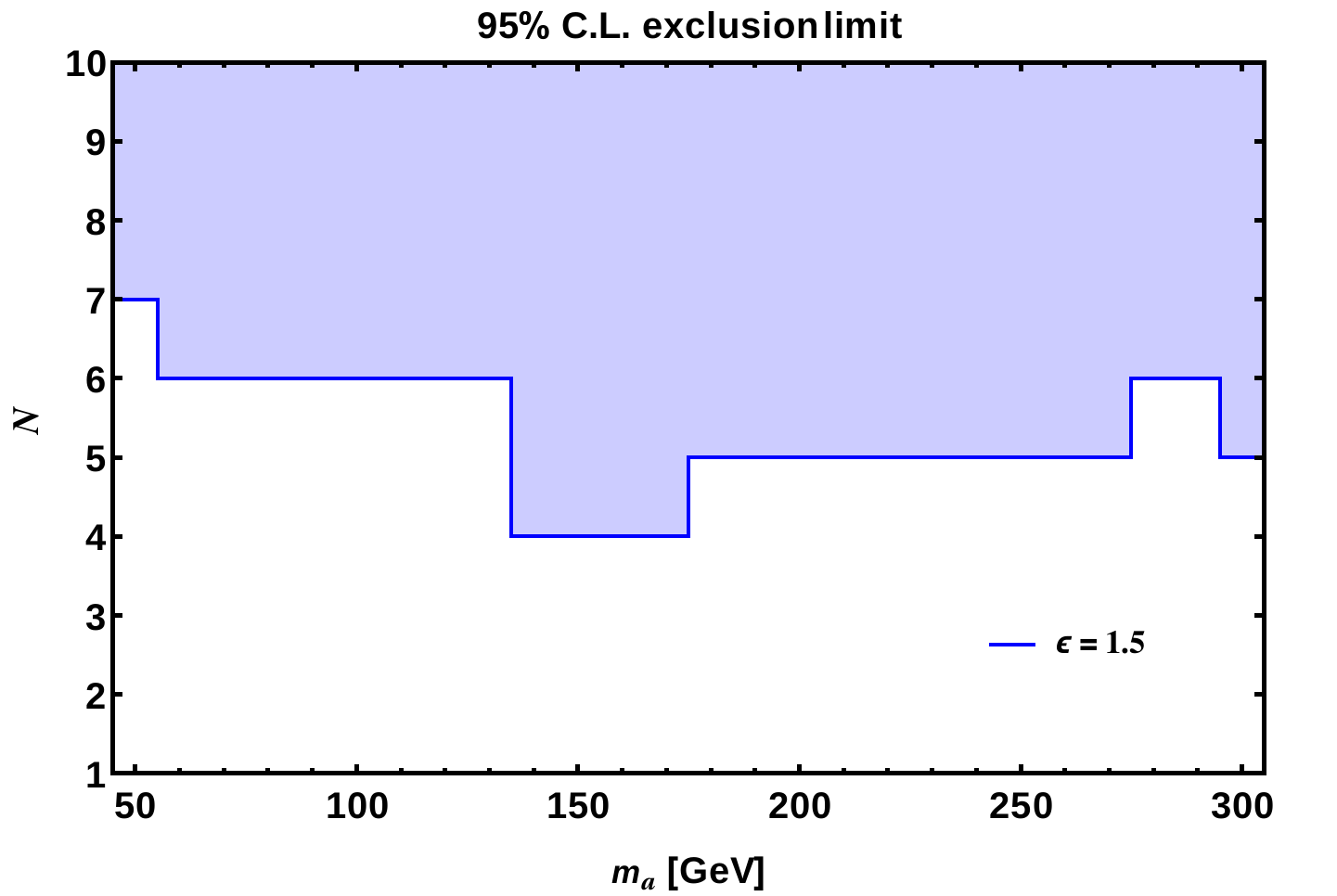} &
\includegraphics[width=83mm]{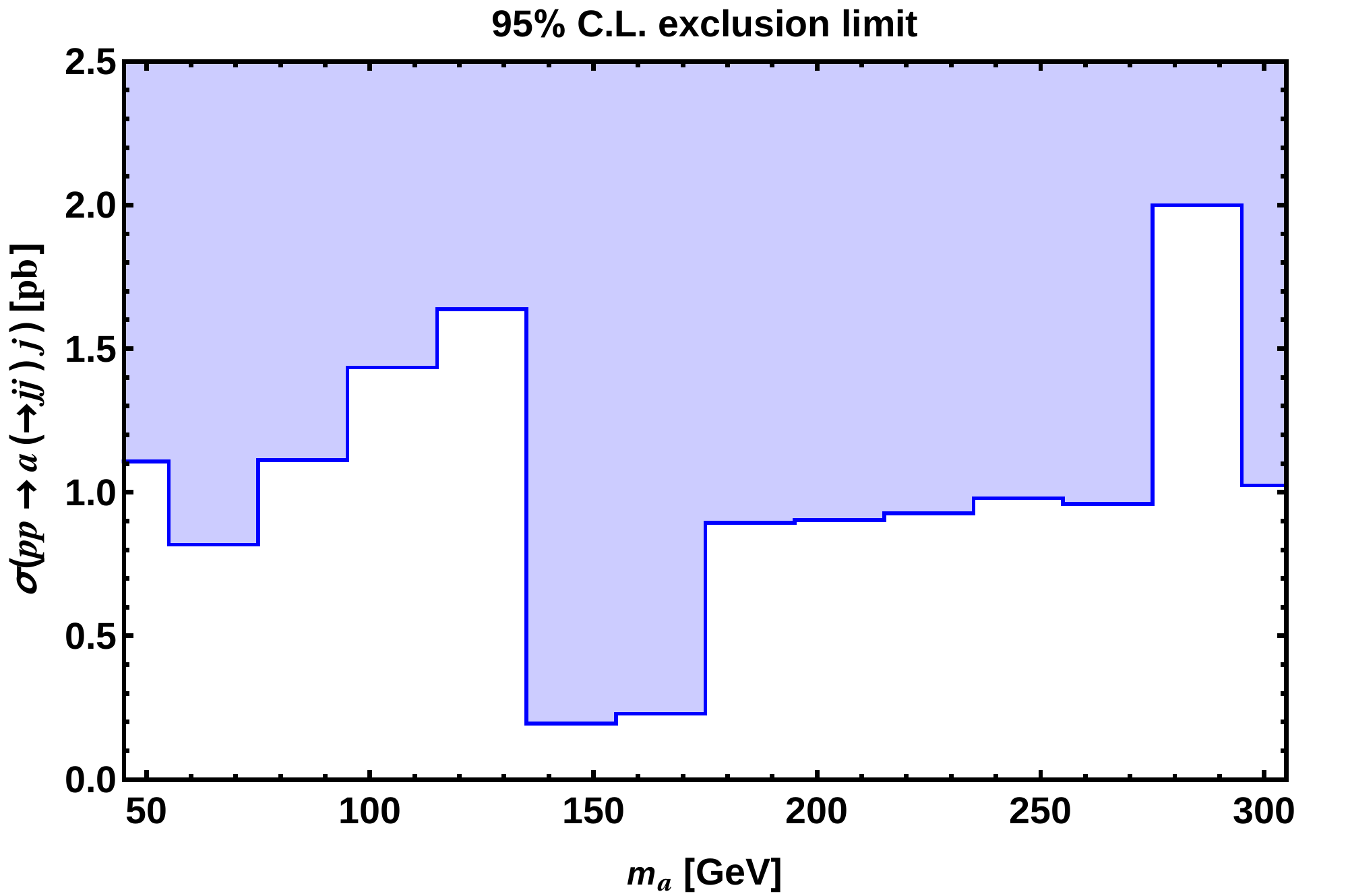}
\end{tabular}
\caption{Left panel: 95\% C.L. exclusion limits for CMS data in the [$m_a,N$] plane. Right panel: 95\% C.L. exclusion limits of the signal cross section for CMS data in terms of $m_a$. This cross section has been computed with {\tt MadGraph 5} with matching up to one extra jet after imposing $p_T^\text{leading-jet} >$ 300 GeV at the generation level and it includes an R-axion mass dependent $K$-factor correction~\cite{Ahmed:2016otz}. In both cases the value of $f_a$ is set to 1 TeV.}
\label{maN_xs-exclusions_CMS}
\end{center} 
\end{figure}

 For purposes of establishing the exclusion limits on the parameter space of the R-axion model, we use the test statistic $q_{\mu}$, which is based on the profile likelihood ratio, and its corresponding $p$-value, denoted as $p_{\mu}$~\cite{Cowan:2010js}. Notice that the five mutually exclusive $p_T$ regions were taken into account in the computation of the likelihood function. The region in the $[m_a,N]$ plane excluded at 95\% C.L. by the CMS data is displayed on the left panel of Fig.~\ref{maN_xs-exclusions_CMS} for $f_a=1$ TeV. We show the same limits translated to the signal cross section on the right panel. As can be seen from this figure, values of the number of color messengers $N$ equal or greater than 7 are forbidden at 95\% C.L. within the considered range for $m_a$. In the region around $m_a =$ 115 GeV, where CMS has reported a slight deviation from the SM expectation, values of $N \geq$ 6 are excluded.
In order to quantify the impact of a potential systematic uncertainty in our estimate of the factor used to correct the invariant mass distributions, we recast the exclusion limits by considering different values of $\epsilon$ both above and below the estimated value (1.5). We found that the exclusion limits are not affected at all by variations of about 10\%, while deviations of up to 30\% shift the contour only in three mass bins out of the 14 considered in the analysis.

\begin{figure}[t!]
\begin{center}
\centering
\includegraphics[scale=0.75]{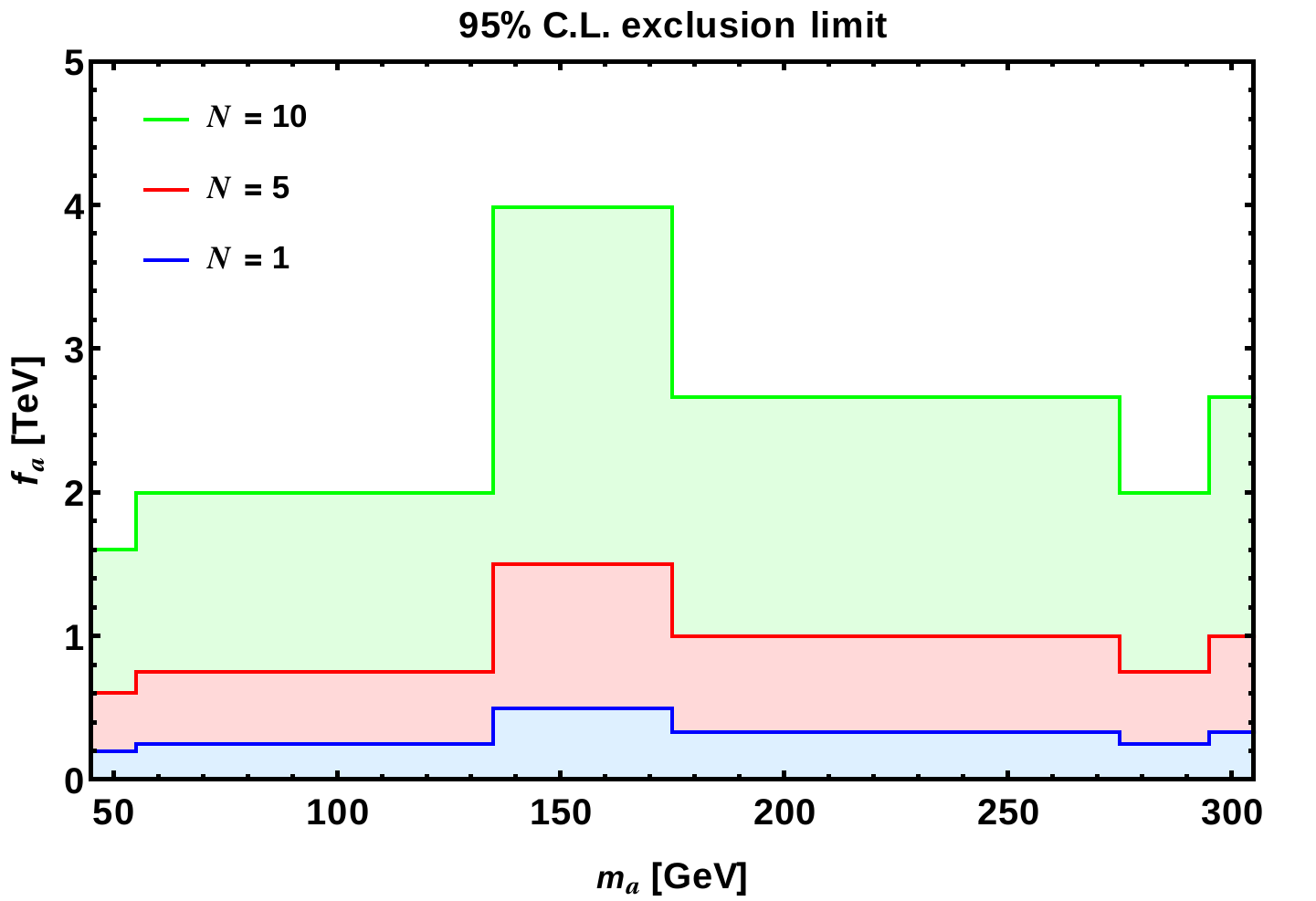}
\caption{95\% C.L. exclusion limits for CMS data in the [$m_a,f_a$] plane for $N=1,\,5,\,10$.}
\label{mafa-exclusions_CMS}
\end{center}
\end{figure}

It is interesting to reinterpret the exclusion limits in the $[m_a,f_a]$ plane with the number of color messengers fixed. In Fig.~\ref{mafa-exclusions_CMS} we show the region excluded at 95\% C.L. by the CMS data for three different values of $N$. In contrast to Fig.~\ref{maN_xs-exclusions_CMS}, now the lower region of the plane $[m_a,f_a]$ is excluded due to the fact that the couplings are inversely proportional to the decay constant. Notice that the exclusion limits become stronger as the value of $N$ increases. In addition, only the range $f_a >$ 4 TeV is allowed for any of the values of $m_a$ and $N$ considered along this work.

\begin{figure}[t!]
\begin{center}
\centering
\includegraphics[scale=0.75]{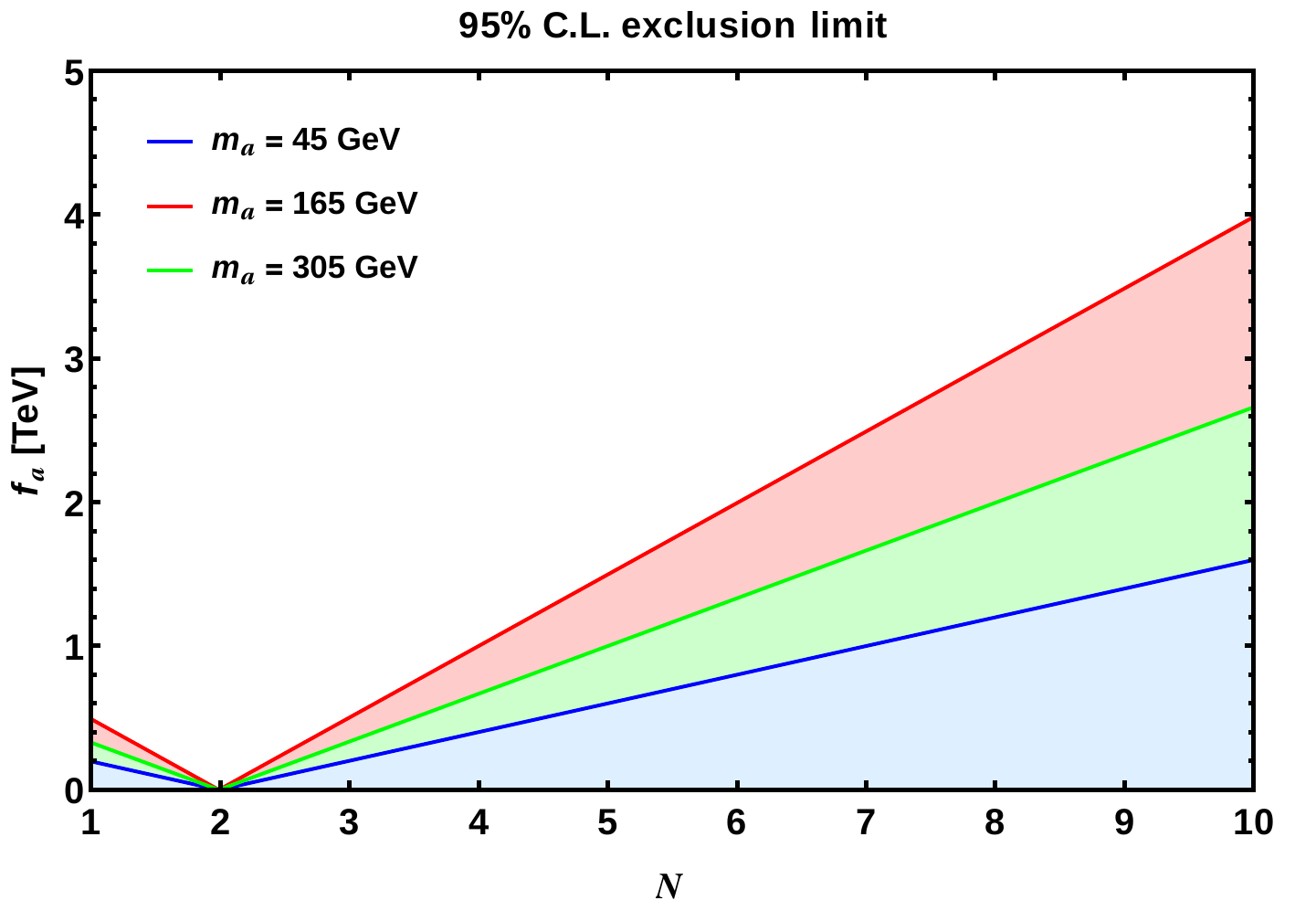}
\caption{95\% C.L. exclusion limits for CMS data in the [$N,f_a$] plane for $m_a=45,\,165,\,305$ GeV.}
\label{Nfa-exclusions_CMS}
\end{center}
\end{figure}

Finally, we display in Fig.~\ref{Nfa-exclusions_CMS} the results in the [$N,f_a$] plane for $m_a=45,\,165,$ and $305$ GeV. Due to the particular suppression that occurs in the coupling of the R-axion to gluons for $N=2$, no constraint can be put in this case and, consequently, all the three exclusion regions depicted in Fig.~\ref{Nfa-exclusions_CMS} collapse to $f_a\simeq 0$ at $N=2$. On the other hand, the linear behavior of the exclusion region boundary obtained in the [$N,f_a$] plane can be read off directly from the dependence of the anomalous coupling to gluons on the parameters $f_a$ and $N$. The fact that the exclusion region for $m_a= 165$ GeV is wider than for $m_a=305$ GeV is not surprising but consistent with what we previously shown in Fig.~\ref{maN_xs-exclusions_CMS}, where it can be seen that the limit set to the signal cross section is more stringent in the former case than in the latter.    

\begin{figure}[t!]
\begin{center}
\centering
\hspace*{-1cm}
\includegraphics[scale=0.53]{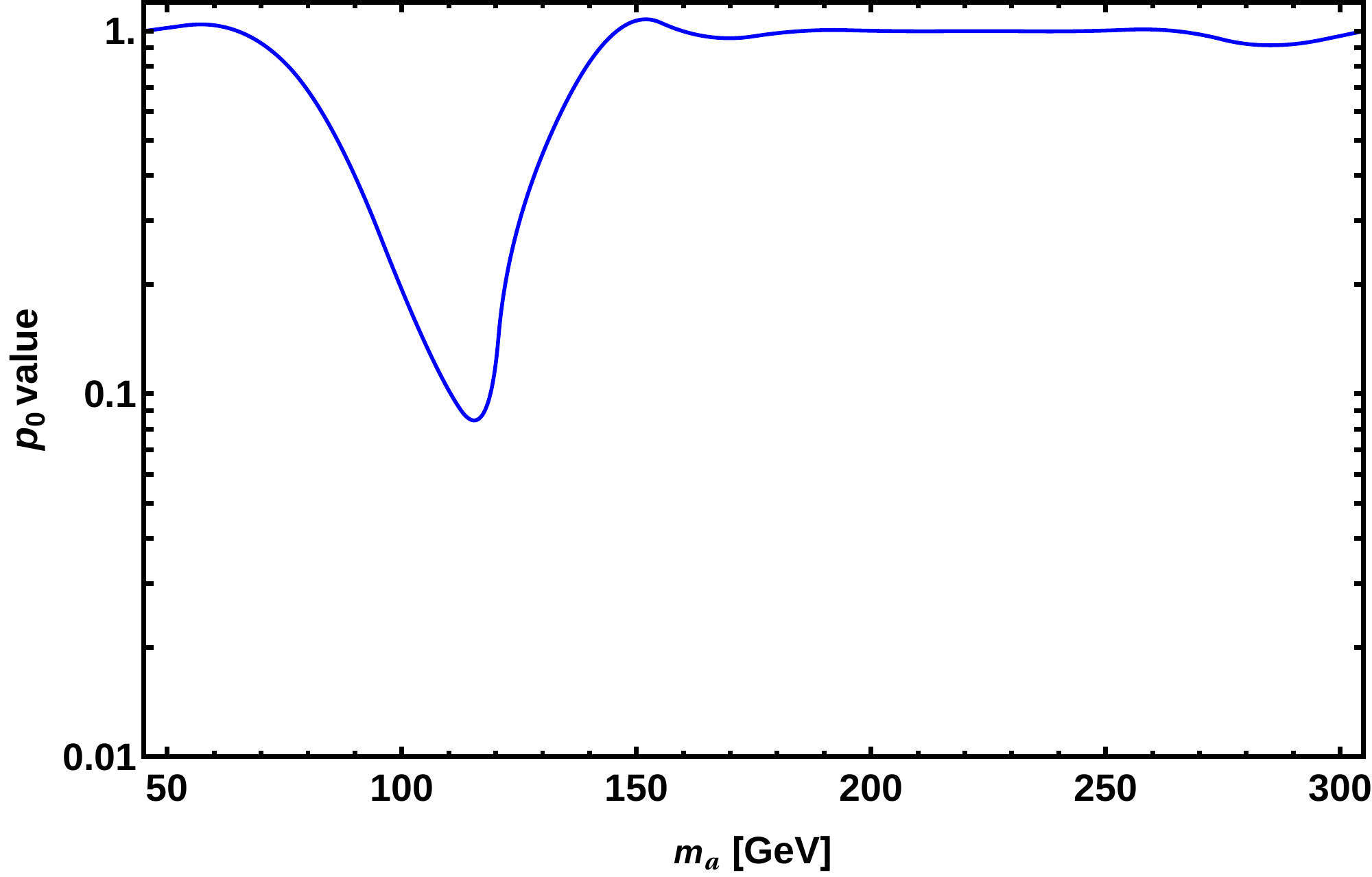}
\caption{$p_0$ value for CMS data as a function of $m_a$ for $f_a=1$ TeV.}
\label{p0-exclusions_CMS}
\end{center}
\end{figure}

As described in Section~\ref{sec:CMS}, CMS has observed slight local and global excesses from the background-only hypothesis at $m_{{\rm SD}}\simeq 115$ GeV. Although the reported values are not statistically significant, we studied the level of agreement between the data and the background-only hypothesis in the case of the R-axion model. To quantify this, we used the statistic $q_0$ and computed the corresponding $p$-value, denoted usually as $p_0$~\cite{Cowan:2010js}. In Fig.~\ref{p0-exclusions_CMS} we display the $p_0$ values as a function of the R-axion mass for the decay constant $f_a=1$ TeV. The minimum value of $p_0$, $\sim 0.09$, is obtained at $m_a=117.5$ GeV ($N=5$) and corresponds to a significance of $\sim 1.35$ standard deviations, which is below the significances obtained within the context of the $Z'$ model considered in the CMS search. This is mainly caused by the discrete nature, through the parameter $N$, of the R-axion model, which constrains the optimization of the likelihood function\footnote{Notice that if one allows free variations of $f_a$, larger significances are possible.}.   

\begin{figure}[t!]
\begin{center}
\centering
\hspace*{-1cm}
\includegraphics[scale=0.53]{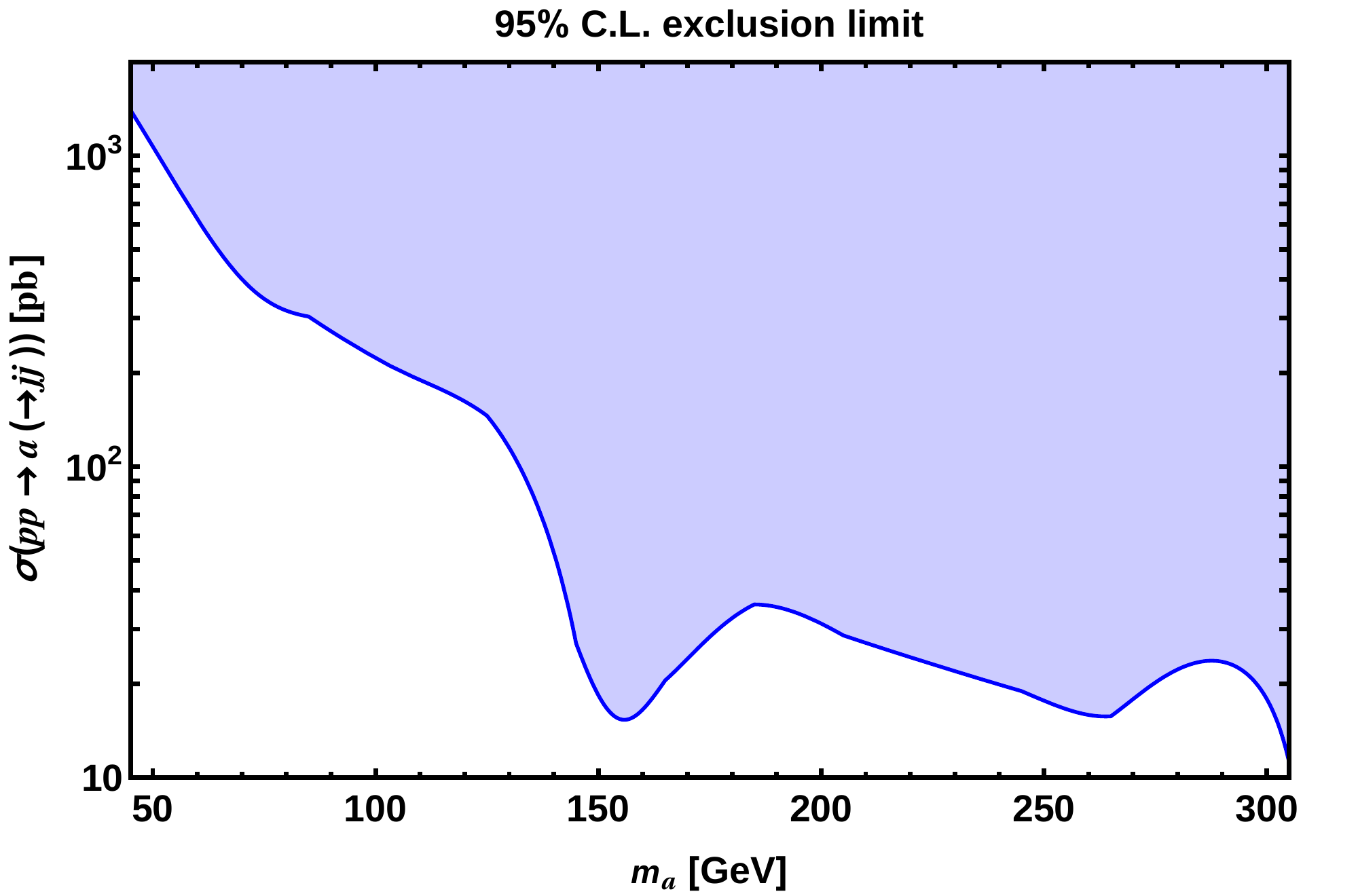}
\caption{95\% C.L. exclusion limits for the gluon-fusion production cross section at the LHC of an ALP decaying into a jet pair as a function of its mass.}
\label{XS-general-exclusions}
\end{center}
\end{figure}

Lastly, we would like to comment that though we worked in the particular context of the R-axion model with a set of R-charges, the constraints derived from the specific CMS collider search can be easily generalized to other ALP models  in which the ALP coupling to photons and SM fermions are suppressed, while the coupling to gluons is enhanced. In fact, this particular CMS search has been considered in \cite{Mariotti:2017vtv} in order to put bounds on ALP models in the mass range between 10 GeV and 65 GeV. As a matter of fact, if one were to consider as main motivation for an axion particle the strong CP problem, there is in principle no reason to expect an anomalous coupling to photons given that there is no CP problem for QED. This coupling to photons is usually induced by the fermion electroweak charges to which the axion couples to. However, there is not necessarily in principle a reason for which these fermions should have EW charges in order to solve this QCD anomalous coupling. In that sense, an ALP like the one we are considering could be related more in the spirit to the strong CP problem.
In Fig.~\ref{XS-general-exclusions} we reinterpret the upper bounds derived previously in terms of the R-axion production cross section via gluon fusion\footnote{This cross section has been computed with {\tt MadGraph} with only basic cuts.}, in such a way that they can be easily generalized to ALP models. For mass values below $\sim$150 GeV, the cross sections allowed by data varies from a few of tens of pb to $\sim 10^3$ pb, whilst for larger mass values, the largest allowed cross section is $\sim$40 pb.

\section{Conclusions}
\label{sec:conclusions}
In this paper we have made use of the results reported by the CMS collaboration in the search for dijet resonances at the LHC with the purpose of imposing constraints on a class of models of dynamical SUSY breaking with a boosted anomalous coupling of the R-axion to gluons and a negligible coupling to photons. From a phenomenological point of view, the most relevant parameters of this R-axion model are the number of color messengers, the mass of the R-axion and its constant decay. In order to establish exclusion limits on these parameters we have used the test statistic $q_{\mu}$ and its corresponding $p$-value $p_{\mu}$, based on the profile likelihood ratio. Setting the decay constant to $f_a =$ 1 TeV, values of the number of color messengers $N$ greater or equal to 7 are forbidden at 95\% C.L. for R-axion masses within the considered range, [45 GeV, 305 GeV]. In the vicinity of $m_a =$ 115 GeV, where CMS has reported a slight excess from the background-only expectation, values of $N \geq$ 6 are excluded. On the other hand, only values of the R-axion decay constant $f_a$ above 4 TeV are allowed by the CMS data for any $N$ and the full $m_a$ range considered here.  
Finally, to quantify the level of agreement between the CMS data and the background-only hypothesis in the case of the R-axion model, we have calculated the corresponding $p$-value $p_0$, obtaining a minimum of $\sim 0.09$ at $m_a =$ 117.5 GeV that corresponds to a $\sim 1.35\sigma$ significance.

Last but not least, it is important to mention that the results obtained with this analysis can be easily generalized to more generic ALP models with a similar coupling structure to the R-axion models studied here.

\section*{Acknowledgments}
We thank Aurelio Juste and Carlos Wagner for useful discussions. E.A. warmly thanks IFT of Madrid for its hospitality hosting him during the completion of this work. This work has been partially supported by CONICET and ANPCyT project no. PICT 2016-0164.


\bibliographystyle{unsrt}

\end{document}